\documentclass[journal]{IEEEtran}
\usepackage{bbm}
\usepackage{amsmath,graphicx,amssymb}
\usepackage{mathtools,dsfont,amsfonts,units,amsthm,bm}
\usepackage{algorithm}
\usepackage{enumerate}
\usepackage[noend]{algpseudocode}
\usepackage{tcolorbox}
\usepackage{graphicx} 
\usepackage{booktabs} 
\usepackage{amsfonts} 
\usepackage{amsthm}
\usepackage{amsmath,graphicx,amssymb}
\usepackage{mathtools,dsfont,amsfonts,units,amsthm,bm}
\usepackage{algorithm}
\usepackage{mdframed}
\usepackage{epstopdf}
\usepackage{subcaption}

\newcommand{\NMSE}{\text{NMSE}}

\newcommand{\x}{\mathbf{x}}
\newcommand{\cb}{\mathbf{c}}
\newcommand{\ellb}{\boldsymbol{\ell}}
\newcommand{\y}{\mathbf{y}}
\newcommand{\fb}{\mathbf{f}}
\newcommand{\A}{\mathbf{A}}
\newcommand{\ab}{\boldsymbol{a}}
\newcommand{\z}{\mathbf{z}}
\newcommand{\Sigmab}{\boldsymbol{\Sigma}}
\newcommand{\omegab}{\boldsymbol{\omega}}

\newcommand{\Oc}{\mathcal{O}}
\newcommand{\Vb}{\mathbf{V}}

\newcommand{\fbh}{\hat{\mathbf{f}}}

\newcommand{\Dc}{\mathcal{D}}
\newcommand{\Sc}{\mathcal{P}}

\newcommand{\E}{\mathbb{E}}
\newcommand{\R}{\mathbb{R}}

\newcommand{\la}{\lambda}
\newcommand{\sig}{\sigma}

\DeclareMathOperator*{\tr}{tr}

\newcommand{\vp}{\vspace{3pt}}
\newcommand{\SNR}{\text{SNR}}
\newcommand{\nn}{\notag}

\newcommand{\ellc}{(\ellb,\cb)}

\newcommand{\cmmnt}[1]{\ignorespaces}

\definecolor{darkred}{RGB}{150,0,0}
\definecolor{darkgreen}{RGB}{0,150,0}
\definecolor{darkblue}{RGB}{0,0,150}
\hypersetup{colorlinks=true, linkcolor=red, citecolor=darkgreen, urlcolor=darkblue}

\begin{document}
\title{
Exploiting Occlusion in\\ Non-Line-of-Sight Active Imaging
}
\author{
Christos~Thrampoulidis$^{1,\ast}$,
Gal~Shulkind$^{1,\ast}$,
Feihu Xu$^{1}$,\\
William~T.~Freeman$^{2,3}$,
Jeffrey~H.~Shapiro$^{1}$,
Antonio~Torralba$^{2}$,
Franco N.~C.~Wong$^{1}$,
Gregory~W.~Wornell$^{1}$,
\thanks{$^\ast$These authors contributed equally.}
	\thanks{This work was supported, in part, by the DARPA REVEAL
          program under Contract No.~HR0011-16-C-0030, and by NSF
          under Grant No.~CCF-1717610.}%
	\thanks{The authors are with the Department of Electrical Engineering
		and Computer Science,
                Massachusetts Institute of Technology,
                Cambridge, MA 02139,
                {$^{1}$Research Laboratory of Electronics}.
{$^{2}$Computer Science and Artificial Intelligence Laboratory}
{$^{3}$Google Research}
.
                 (E-mails: $\{$cthrampo, shulkind, fhxu, billf, jhs, torralba, ncw, gww$\}$@mit.edu.)}}

\maketitle
\begin{abstract}
Active non-line-of-sight imaging systems are of growing interest for
diverse applications.  The most commonly proposed approaches to date
rely on exploiting time-resolved measurements, i.e., measuring the
time it takes for short light pulses to transit the scene. This
typically requires expensive, specialized, ultrafast lasers and
detectors that must be carefully calibrated.  We develop an
alternative approach that exploits the valuable role
that natural occluders in a scene play in enabling accurate and
practical image formation in such settings without such hardware
complexity.  In particular, we demonstrate that the presence of
occluders in the hidden scene can obviate the need for collecting
time-resolved measurements, and develop an accompanying analysis for
such systems and their generalizations.  Ultimately, the
results suggest the potential to develop increasingly sophisticated
future systems that are able to identify and exploit diverse
structural features of the environment to reconstruct scenes hidden
from view.
\end{abstract}

\begin{IEEEkeywords}
computational imaging, computer vision, non-line-of-sight imaging,
time-of-flight cameras, LIDAR
\end{IEEEkeywords}

\section{Introduction}


In contrast to classical photography, where the scene of interest is
in the observer's direct line of sight, non-line-of-sight (NLOS)
imaging systems only have indirect access to a scene of interest via
reflections from intermediary surfaces.  Such systems are of
considerable interest for applications spanning a wide variety of
fields including medicine, manufacturing, transportation, public
safety, and basic science.

Despite their obvious appeal, there are inherent challenges in the
design of NLOS systems.  In particular, typical surfaces (e.g., walls,
floors, etc) diffusely reflect light, effectively removing beam
orientation information, and rendering the problem of scene
reconstruction poorly conditioned.  In order to compensate for the
losses induced by diffuse reflections, initial demonstrations of NLOS
imaging used ultrafast transient-imaging modalities
\cite{kirmani2011looking,velten2012recovering} that involved a laser
source to send optical pulses of sub-picosecond duration, and a streak
camera exhibiting temporal resolution in the picosecond range. A
computational algorithm then used the fine time-resolved light
intensity measurements to form a three-dimensional reconstruction of
the hidden scene. 

The system requirements posed by these systems, for transmission of
very short, high power optical pulses on the transmitter side, and for
very high temporal resolution on the receiver side, inevitably imply
high system complexity and cost.  Thus, much of the follow-up work has
focused on developing reduced cost and power implementations.  For
example, \cite{buttafava2015non} uses a single-pixel, single-photon
avalance diode (SPAD) detector for reduced power consumption and cost;
\cite{gariepy2015detection} uses a multi-pixel SPAD camera to
demonstrate tracking of hidden moving objects; and
\cite{heide2014diffuse} uses modulated illumination and CMOS
time-of-flight sensors, including photonic mixer devices, to
substantially reduce overall system cost, albeit at the expense of
reduced spatial resolution.

\subsection{Our Contribution}

To address the limitations of such existing approaches, we introduce a
rather different imaging modality for such problems.  In particular,
we develop the beneficial role that natural occlusions---which would
traditionally be viewed as an impediment to imaging---play in actually
facilitating robust image reconstruction in NLOS settings.  In fact,
we demonstrate---analytically and experimentally---that in some cases
the presence of occluders in the hidden scene can obviate the need for
collecting time-resolved (TR) measurements, enabling imaging systems
of significantly reduced cost.  In turn, and in contrast to existing
methods, this means 
our approach is compatible with wide field-of-view
detectors, enabling the collection of more photons per measurement and
accelerating acquisition times so as to facilitate real-time
operation.

We introduce the key concepts and principles in the context of imaging
a hidden wall of unknown reflectivity.  For this problem, we develop a
framework of analysis that involves a mathematical formulation, as
well as numerical and experimental illustrations. We further study
diverse features of the proposed occlusion-based imaging system, such
as robustness to modeling errors, and optimal selection of
measurements. More generally, the ideas that we introduce open
opportunities in designing more accurate, robust, and cost-effective
NLOS imaging systems that relax the stringent temporal resolution
requirements for optical measurements in the presence of occluders.
We envision that this motivates further research towards the
development of NLOS imaging systems that opportunistically exploit
known structural features in the environment, such as occluders.

\subsection{Related Work}

To the best of our knowledge, this paper is the first to develop the
role of exploiting occluders for NLOS imaging.  However, there is a
variety of related work in computational imaging more generally that
investigates exploiting physical structure in the space between the
scene of interest and the measurement system.  Perhaps the best known
is that on what is referred to as ``coded-aperture imaging,'' in which
occlusion in the optical path takes the form of a carefully designed
physical mask that modulates the light transferred from the scene of
interest to a detector array.  Among the earliest and simplest
instances of coded-aperture imaging are those based on pinhole
structure \cite{fenimore1978coded} and pinspeck (anti-pinhole)
structure \cite{cohen1982anti}, though more complex structure is
commonly used.  Such methods are of particular interest in
applications where lens fabrication is infeasible or impractical, such
as in x-ray and gamma-ray imaging.  More generally, a number of rich
extensions to the basic methodology have been developed; see, e.g.,
\cite{brady2004reference} and the references therein.

In other developments, the value of using a mask in conjunction with a
lens has been investigated in computational photography for motion
deblurring \cite{raskar2006coded}, depth estimation
\cite{levin2007image}, and digital refocusing and recovery of 4D
light-fields \cite{veeraraghavan2007dappled}.  More recently, there
has been an increased interest in using masks with appropriate
computational techniques, instead of traditional lens-based cameras,
to build cameras that have fewer pixels, need not be focused
\cite{duarte2008single}, and/or meet physical constraints
\cite{asif2015flatcam}.  All these methods are passive imagers; only
very recently has the addition of an active illumination source and
time-resolved sensing been proposed to reduce acquisition time
in lensless systems \cite{satat2016lensless}.

Perhaps the work most closely related to the present paper that
demonstrating how information about a scene outside the direct field
of view can be revealed via ``accidental'' pinhole and anti-pinhole
camera images \cite{torralba2012accidental}.  The accidental camera is
based on the use of video sequences obtained only with ambient
illumination, and requires a reference frame without the occluder
present.  In effect, the present paper can be viewed as quantifying
the higher-resolution imaging performance achievable without these
limitations, and in particular when we actively illuminate the scene
with a scanning laser.

Finally, there have been recently demonstrations of a method for
tracking specifically moving objects in NLOS scenes via
non-time-resolved intensity images of a visible wall
\cite{klein2016tracking}.  By contrast, our framework emphasizes
imaging without requiring the presence---and exploitation---of scene
motion, so can be applied much more broadly.

\subsection{Paper Organization}

The paper is organized as follows.  Section \ref{sec:fwd_model}
introduces a forward propagation model for NLOS imaging that accounts
for sources of occlusion, and Section \ref{sec:framework} introduces a
framework of analysis for NLOS imaging in the presence of such
occlusion.  Section \ref{sec:tr_reconstruction} then establishes the
limitations of time-resolved measurements with respect to the temporal
resolution of the detector, and Section \ref{sec:occ_reconstruct}
shows how to transcend these limitations by opportunistically
exploiting occluders in the hidden scene.  An experimental
demonstration of the methodology is presented in Section
\ref{sec:experimental_results}.  Finally, Section \ref{sec:discussion}
contains a discussion of extensions and
opportunities for future research.

\section{Forward model for NLOS Imaging}\label{sec:fwd_model}

\begin{figure}[t]
	{\includegraphics[width=0.8\columnwidth]{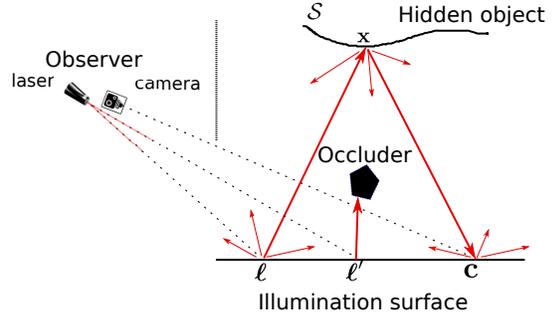}}
	\caption{
	Red lines trace beam paths reflecting from the virtual laser points $\ellb, \ellb'$, where a laser beam hits the illumination \cmmnt{wall} surface towards point $\x$ on the hidden object. The beam emanating from $\ellb'$ is blocked by the occluder. 
	Upon hitting the point $\x$ light reflects back towards a virtual camera position $\cb$, where a focused camera is steered.} \label{fig:5d_setup}
\end{figure}

The goal of NLOS imaging systems is to process reflected
light-intensity measurements and perform joint estimation of both the
geometry and reflectivity properties of a hidden three-dimensional
scene, as illustrated in Figure \ref{fig:5d_setup}.  A focused laser
beam is steered towards a visible \emph{illumination} surface
\cmmnt{(or, visible wall)} and reflects back towards a hidden
\cmmnt{scene} object. Upon hitting the object light is reflected back
towards the illumination surface and is measured by a focused camera.
This forms a three-bounce problem in which light beams follow paths of
the
form 
\begin{equation*}
\text{Laser} \rightarrow \ellb \rightarrow \x \rightarrow \cb
\rightarrow \text{Camera},   
\end{equation*}
where $\ellb,\cb$ lie on the illumination surface \cmmnt{wall}
\cmmnt{(or \emph{reflectance wall})} and $\x$ lies on the hidden
object surface. By raster scanning the laser and/or changing the focal
point of the camera, we retrieve multiple measurements corresponding
to a set of $K$ parameters $\Sc=\left\lbrace(\ellb_i,\cb_i)\vert
i=1,\ldots,K\right\rbrace$.

In this section we formulate a forward propagation model that determines the irradiance waveform $y_{\ellb,\cb}(t)$ measured at point $\cb$ on the illumination surface in response to a single optical laser pulse $p(t)$ fired towards position $\ellb$. We let $\mathcal{S}$ be a parametrization of the hidden object surface, and $f(\x), \;\x\in \mathcal{S}$ denote the spatially varying reflectivity function (or, albedo). The model assumes that the illumination and hidden object surfaces are both ideal Lambertian reflectors. 

In order to account for the presence of occluders in the scene (as illustrated in Figure \ref{fig:5d_setup}), we introduce a binary \emph{visibility function} $V(\x,\z)$ which determines whether point $\x$ on the hidden object surface $\mathcal{S}$ and point $\z$ on the illumination surface are visible to each other:
\begin{align}\label{eq:vis}
V(\x,\z)=\left\lbrace
\begin{array}{ll}
1, & \text{clear line of sight between $\x$ and $\z$,}\\
0, & \text{no line of sight between $\x$ and $\z$.}
\end{array}
\right.
\end{align}
With these, the forward model is given as follows\footnote{A similar forward model is used in \cite{heide2014diffuse}, and is based on well-known principles, namely quadratically decaying power attenuation for optical beams, and Lambert's cosine law for diffuse reflection. Eqn. \eqref{eq:fwd_model} further accounts for possible occlusions in the scene through the visibility function.}:
\begin{align}\label{eq:fwd_model}
&y_{\ellb,\cb}(t)=\int_{{\mathcal{S}}} f(\x)
\frac{V(\x,\ellb)V(\x,\cb)}{\|\x-\ellb \|^2\|\x-\cb\|^2}G(\x,\ellb,\cb)\cdot\nn
\\&\qquad\qquad\qquad
\cdot p\!\left(t-\frac{\|\x-\ellb \|+\|\x-\cb\|}{c}\right){\rm d}\x.
\end{align}
Here,  $G$ is the Lambertian Bidirectional Reflectance Distribution Function (BRDF):
\begin{align*}
G(\x,\ellb,\cb)
\equiv&\cos(\x-\ellb,\mathbf{n}_{\ellb})\cos(\x-\ellb,\mathbf{n}_\x)\cdot\nn
\\
&\quad\quad\cdot\cos(\x-\cb,\mathbf{n}_{\x})\cos(\x-\cb,\mathbf{n}_\cb),\nn
\end{align*}
$\mathbf{n}_\x,\mathbf{n}_\cb,\mathbf{n}_{\ellb}$ are the surface normals at $\x, \cb, \ellb$, respectively and $c$ is the speed of light.
The model can easily be generalized to account for non-Lambertian BRDFs for the illumination wall and the hidden object by appropriately adjusting $G$. 

Several remarks are in place with respect to \eqref{eq:fwd_model}:

\paragraph*{Virtual laser and camera positions}
For simplicity in the exposition we have excluded from the model the attenuation, delay, and BRDF contributions accrued along the path from the laser to $\ellb$ as well as those accrued from $\cb$ to the camera. Note that those quantities are fixed and known to the observer, hence can be easily compensated for.
In general, it is useful for our exposition to think of $\ellb$ and
$\cb$ as \emph{virtual unfocused illumination and camera positions}
(Figure \ref{fig:5d_setup}), and Equation \eqref{eq:fwd_model} is
consistent with that. 

\paragraph*{Visibility function} The visibility function in
\eqref{eq:fwd_model} accounts for obstructions of light beams in the
imaging process, tracking hidden object patches that are either not
reached by the virtual illumination at $\ellb$ or are not observable
from the virtual camera at $\cb$. Implicit in this description is the
partition of the objects occupying the space facing the illumination
wall in: (a) the \textit{hidden objects}, which are objects of
interest in the 
reconstruction process; (b) the \textit{occluders}, which are not of
immediate interest (in fact, we usually assume that they are
known), 
blocking at least some light beams between the illumination and hidden
object surfaces. 

\paragraph*{Third-bounces} The model \eqref{eq:fwd_model} accounts for
the contributions in the measurements resulting from three bounces (at
$\ellb,\x,\cb$) that are informative about the hidden
objects. Higher-order bounces are neglected, since they typically
experience high attenuation in the setting considered.  
Also, in deriving  \eqref{eq:fwd_model} we model the
occluders as fully absorbing objects\footnote{This model also applies
  for reflective occluders of known reflectivity pattern since their
  contribution in the measurements can be compensated for.}. 

\paragraph*{Temporal resolution of the camera}   The camera averages
the incident irradiance at $\cb$ with a finite temporal resolution
$\Delta t$ resulting in measurements $\y_{\ellb,\cb,\tau},
\tau=1,2,\ldots,T$,  
\begin{align} \label{eq:finite_res_samples}
\y_{\ellb,\cb,\tau} = \int_{(\tau-1) \Delta t}^{\tau\Delta t}
\y_{\ellb,\cb}(t) \mathrm{d}t.
\end{align}
Since only third-bounce reflections involving the hidden object are of
interest to us, with some abuse of notation we shift the time axis
such that time $t=0$ is the first instant when third bounce
reflections reach the camera and $T \Delta t$ is chosen such that all
relevant third-bounce reflections from the hidden object are included
in the interval $[0,T \Delta t]$.

\begin{figure}[t!]
	\centering
	\includegraphics[width=.6\columnwidth]{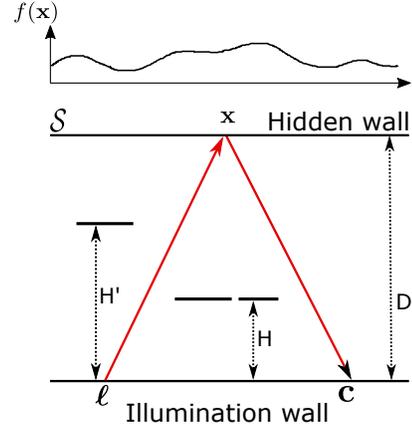}
	\caption{ The proposed \cmmnt{scene model} imaging setting in which the objective is to reconstruct the reflectivity $f(\x)$ of a flat hidden wall that is parallel to the illumination wall at known distance $D$. 
	The position and size of the fully absorbing occluders are known.}
	 \label{fig:study_setup}
\end{figure}

\section{Scene and system model}\label{sec:framework}

To develop the key principles 
of analysis, we turn to a specific
instance of the general NLOS imaging scenario described in the
previous section (also, Figure \ref{fig:5d_setup}), which we now
describe. Extensions are discussed in Section \ref{sec:discussion}.

\subsection{A Representative Imaging Setting}\label{sec:setup}
Our 
setup is illustrated in Figure~\ref{fig:study_setup}. It includes a planar hidden object and a parallel planar illumination surface, which we refer to as the \emph{hidden wall} and the \emph{illumination wall}, respectively. These two surfaces of known geometry 
are placed distance $D$ apart. 
In between the illumination and the hidden walls lie occluders, whose effect on the imaging process is captured through the visibility function defined in \eqref{eq:vis}. The occluders are fully absorbing objects of known geometry and location. Hence, the visibility function is known. The NLOS imaging objective under this setting is then to reconstruct the unknown reflectivity function $f(\x)$ of the hidden wall from the measurements.
%


Under the aforementioned setting, the measurements $\y_{\ellb,\cb,\tau}$ in \eqref{eq:finite_res_samples} are \emph{linear} in the unknown reflectivity function $f(\x)$. Let $\x_1,\ldots,\x_N$ be a discretization of the hidden wall, then, according to \eqref{eq:fwd_model}, each measurement $\y_{\ellb,\cb,\tau}$ corresponds to a measurement vector $\ab_{\ellb,\cb,\tau}\in\R^N$ such that $\y_{\ellb,\cb,\tau} = \ab_{\ellb,\cb,\tau}^\top\fb$, where $\fb := [f(\x_1), \ldots, f(\x_N)]^\top$.
Repeating the measurements for a total of $K$ $\ellc$ pairs, obtaining $T$ time samples per each pair, and collecting them in a vector $\y$ of dimension $M=K\cdot T$, this gives rise to the linear system of equations $\y=\A\fb$ where $\A$ is an $M\times N$ measurement matrix whose rows are vectors $\ab_{\ellb,\cb,\tau}^\top$ that correspond to the chosen $\ellc$ pairs and temporal resolution $\Delta t$. In this study we consider measurements that are contaminated by additive noise $\boldsymbol{\epsilon}$:
\begin{align}\label{eq:noisy_lin}
\y = \A\fb + \boldsymbol{\epsilon}.
\end{align}
The noise term can be thought of as a simple means to capture system modeling  errors, camera quantization errors, background noise, etc.. We study generalizations to other noise models in \cite{experiment}.

\subsection{Bayesian Scene Model}\label{sec:bayes}
The idea of imposing Bayesian priors is well-established in image processing \cite{besag1991bayesian,geman1984stochastic}: past studies have considered various forms of Gaussian prior distributions on the unknown target scene, including variations promoting sparse derivatives \cite{levin2007image}, and natural image statistics \cite{mihcak1999low}. 
Such priors offer enough flexibility and at the same time are amenable to analysis and intuitive interpretation.
In this work, we
let\footnote{The zero mean assumption is somewhat simplified, but not particularly restrictive. Strictly speaking, in order to respect the nonnegative nature of the reflectivity function, a positive additive mean should be added in all the scenarios considered in this paper, but this addition has no effect on the qualitative conclusions drawn from our results.}:
\begin{align}\label{eq:bayes_f}
\fb\sim \mathcal{N}(\mathbf{0},\Sigma_\fb),
\end{align}
with a smoothness promoting kernel function such that the entries of the covariance matrix are  $[\Sigma_\fb]_{ij} = \exp(-\frac{1}{2\pi\sig_\fb^2}\|\x_i-\x_j\|^2)$
and the spatial variance $\sig_\fb^2$ controls the extent of smoothness.
Additionally, we consider an i.i.d. Gaussian distribution
for the measurement noise $\boldsymbol{\epsilon}_i\sim\mathcal{N}(0,\sigma^2)$ such that the Signal to Noise Ratio in our problem is given by 
$\SNR = \tr\bigl(\A\Sigmab_\fb\A^\top\bigr)/(M\sig^2)$,
where $M$ denotes the total number of measurements.
For the  reconstruction, we consider the minimum mean-squared error (MMSE) estimator, which under the Gaussian framework is explicitly computable as
\begin{align}\label{eq:f_hat}
\fbh = \Sigmab_{\fb}\A^\top( \A\Sigmab_{\fb}\A^\top + \sigma^2\mathbf{I} )^{-1}\y.
\end{align}
We measure and compare reconstruction performance in different settings using the normalized mean squared error $\NMSE = {\E\|\fbh-\fb\|_2^2}/{\E\|\fb\|_2^2},$ which equals the (normalized) trace of the posterior covariance matrix
\begin{equation*}
\NMSE = \frac{1}{M}\tr( \Sigmab_{\fb} - \Sigmab_{\fb}\A^\top(
\A\Sigmab_{\fb}\A^\top + \sigma^2\mathbf{I} )^{-1}\A\Sigmab_{\fb} ).
\end{equation*}
Note that the NMSE can be evaluated before collecting measurements $\y$. Also, the reconstruction in \eqref{eq:f_hat} remains the optimal linear estimator under given first and second order statistics for $\fb$, even beyond Gaussian priors.


%
\section{Time-Resolved Measurements} \label{sec:tr_reconstruction}

In this section we study the limits of traditional NLOS imaging that is based on collecting fine time-resolved (TR) measurements, and set up a reference against which we want to compare the newly proposed imaging modality that uses occlusions and no TR information, which we formally introduce in Section \ref{sec:occ_reconstruct}.
%
%

\subsection{Virtues of Time-Resolved Measurements}\label{sec:TR_virtues}
%


Assuming an ideal pulse $p(t)=\delta(t)$, and considering the propagation of optical pulses at the speed of light $c$, the measurement $\y_{\ellb,\cb,\tau}$ taken at time step $\tau$ forms a linear combination of the reflectivity values of only those scene patches $\x_i$ whose sum distance to $\ellb$ and $\cb$ corresponds to a propagation time around $\tau \Delta t $. These patches fall within the elliptical annulus with focal points $\ellb$ and $\cb$ described by the following equation:
$$(\tau-1)\cdot c\Delta t\leq \|\x_i-\ellb\| + \|\x_i-\cb\| \leq \tau\cdot c\Delta t.$$
The thinner the annulus (eqv. the lower $\Delta t$), the more informative the measurements are about the reflectivity values of these patches. Furthermore, scanning the laser and camera positions $(\ellb,\cb)$, different sets of light paths are probed, each generating a different set of elliptical annuli. For a total of $K$ $(\ellb,\cb)$-pairs, this forms the linear system of equations \eqref{eq:noisy_lin}.


\begin{figure}[t]
	\centering
	\makebox[\columnwidth]{\includegraphics[trim={3.5cm 0 2.5cm 0},clip,width=1\columnwidth]{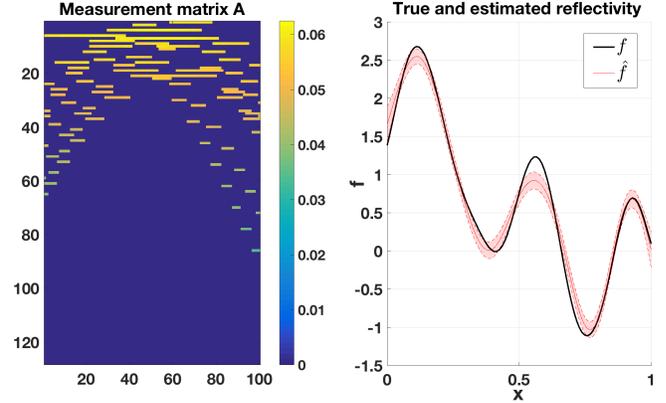}}
	\caption{Reflectivity reconstruction from TR measurements. (left) Measurement matrix, where each row corresponds to a specific choice for the $(\ellb,\cb)$ pair and time index $\tau$. The columns correspond to a discretization of the hidden wall to $N=100$ points. (right) True reflectivity function versus the MMSE estimate $\hat\fb$.
	} \label{fig:time_resolved_reconstruction}
\end{figure}

We performed a simple numerical simulation to demonstrate 
scene reconstruction performance in a TR setup. 
For the purposes of illustration the simulations presented here and in the rest of the paper are in a two-dimensional world. This allows for easy visualization of important concepts such as the visibility function and the forward measurement operator, and it enables useful insights, but is otherwise non-restrictive. 
The room size was set such that the width of the walls is $1$m, the distance between the walls is $D=2$m and the temporal resolution was set at $\Delta t=100 \text{ps}$. $K=8$ $(\ellb,\cb)$ pairs were randomly chosen, $f$ was drawn according to the Gaussian prior with $\sigma_f^2=0.1$, and we set $\text{SNR}=13.7 \text{dB}$. The results are summarized in Figure \ref{fig:time_resolved_reconstruction}, where we plot the measurement matrix $\A$, the true reflectance $\fb$ and the estimated $\fbh$ with the corresponding reconstruction uncertainty depicted in shaded color around the MMSE estimator.
The reconstruction uncertainty for our purposes is the square-root of the diagonal entries in the posterior covariance matrix corresponding to the standard deviation of ${\hat\fb}_i-\fb_i$ for the individual points $i$ on the wall. For this setup and resolution we collect $T=16$ temporal samples per $(\ellb,\cb)$ pair such that the total number of measurements is $M=8\cdot16=128$. These are the rows of $\A$ depicted in the figure, where each block of $8$ consecutive rows corresponds to the measurements collected at a single time instant and for all $(\ellb,\cb)$ pairs. Notice that the last few blocks are zero as at that time no patch on the hidden wall contributes to the measurements any more.

\subsection{Performance Dependence on Temporal-Resolution} \label{sec:TR_res}
%
The simulation results shown in Figure \ref{fig:time_resolved_reconstruction} demonstrate high-fidelity reflectivity reconstruction when the  available temporal resolution is fine ($\Delta t=100 \text{ps}$). However, practical technological and budget considerations limit the availability of such fine TR information and this results  in significant deterioration of the reconstruction fidelity, as we show next. 


Let us first consider an extreme situation where the temporal resolution is so low such that the distance that light travels during a single resolution window of the detector is larger than the entire spatial extent of the scene.
As an example, for the setup in Figure \ref{fig:time_resolved_reconstruction} this happens when $\Delta t\gtrsim 1.5$ ns. In this extreme, which is essentially equivalent to collecting \emph{non-time-resolved measurements}, 
each $(\ellb,\cb)$-pair effectively generates just a single \emph{scalar} measurement which we denote $\y_{\ellb,\cb}$ and which
%
%
 is a linear combination of \emph{all} the entries of $\fb$. The combination coefficients are determined by the decay and cosine factors in \eqref{eq:fwd_model}. Focusing on the distance factors $\|\x-\ellb\|^{-2}\|\x-\cb\|^{-2}$ for intuition, the range of values that these can take is clearly determined by the geometry of the problem, and can be very limited; for example, if the two walls are far apart. This weak variation can result in poor conditioning of the measurement matrix $\A$ and subsequently poor reconstruction fidelity. 

\begin{figure}[t!]
    \centering
    \begin{subfigure}[b]{1\columnwidth}
    	\includegraphics[width=1\textwidth]{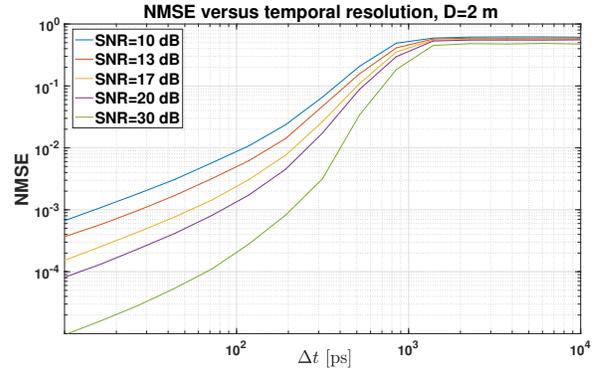}
    	\caption{Normalized mean-squared error in reconstruction versus temporal resolution.} \label{fig:mse_vs_res_vs_SNR}
    \end{subfigure}	
    \quad 
    \begin{subfigure}[b]{1\columnwidth}
	\includegraphics[width=\textwidth]{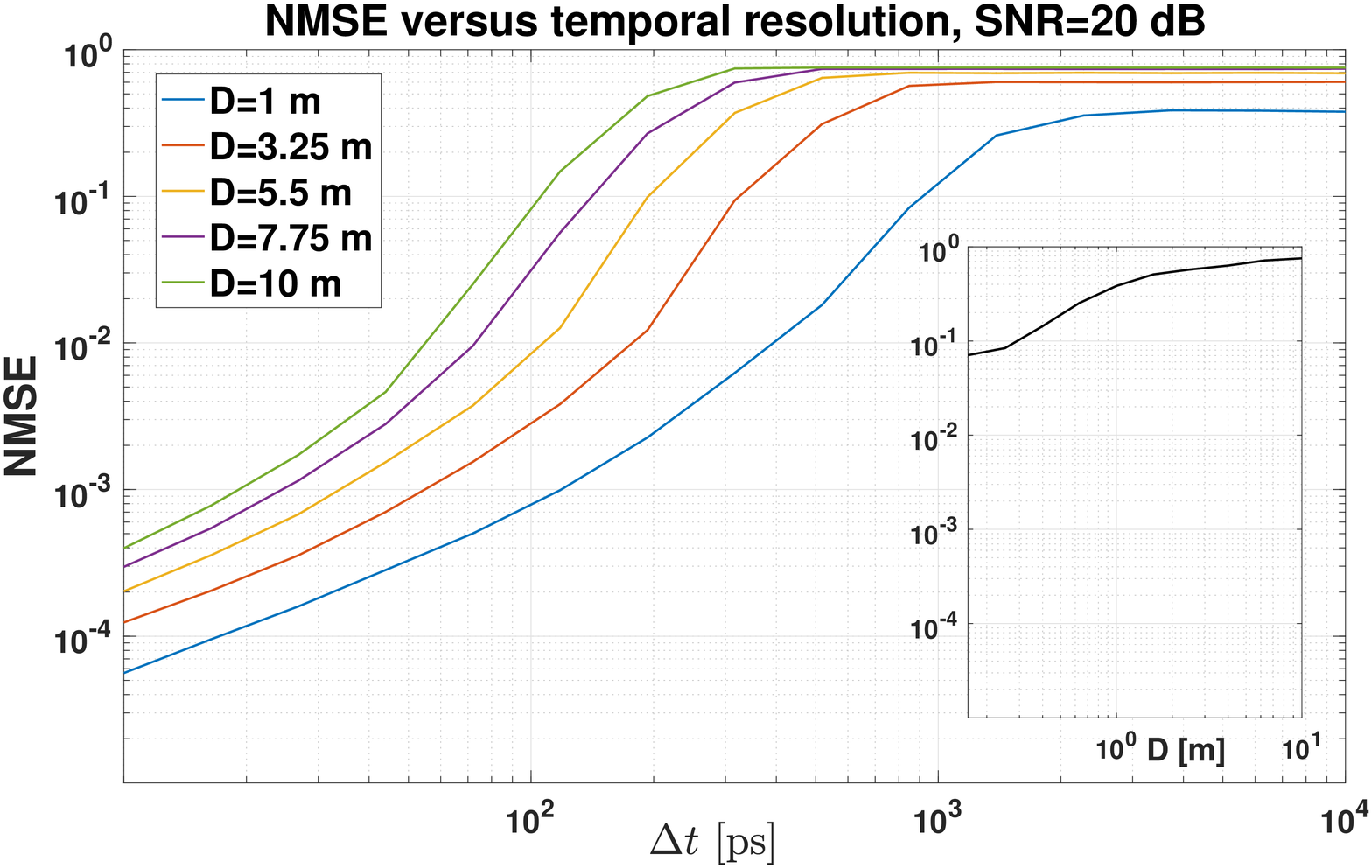}
	\caption{Normalized mean-squared error in reconstruction versus temporal resolution.} 			\label{fig:mse_vs_res_vs_D}
	\end{subfigure}
    \caption{Study of the reconstruction error as a function of the available temporal resolution $\Delta t$ of the detector in TR sensing.} \label{fig:occluder_setup}
\end{figure}

This ill-conditioning is illustrated in Figure \ref{fig:mse_vs_res_vs_SNR} where we plot the NMSE for the same setup as in Fig.~\ref{fig:time_resolved_reconstruction}, with $K=30$ measurements versus the temporal resolution parametrized against the SNR (for each data point
 we average over 10 random draws for $(\ellb,\cb)$). Observe that as $\Delta t$ deteriorates, reconstruction fidelity decreases. Considering finite SNR for the purpose of this evaluation is key as reconstruction in an ideal noise-free experiment could result in high fidelity reconstruction even if $\A$ is ill conditioned\footnote{Each of the plots in Figure \ref{fig:mse_vs_res_vs_SNR} correspond to different $\text{SNR}$. In practice, when comparing setups of different temporal resolving capabilities the equipment involved will be technologically different such that a fair comparison does not necessarily entail assuming a fixed SNR common to all setups. Notice however the general trend of worsening reconstruction performance with diminishing temporal resolutions, which holds for all SNR levels. }. 


When imaging more distant walls, the poor conditioning of $\A$ further deteriorates as the distance decay factors become less varied and approach constants $\|\x-\ellb\|\approx\|\x-\cb\|\approx D$, as illustrated in Figure \ref{fig:mse_vs_res_vs_D} where reconstruction performance is parametrized against $D$ for a fixed $\text{SNR}$ in a setup with otherwise identical parameters as those of the first subfigure.
In particular notice in this plot the limit of non-time-resolved measurements $\Delta t > 1.5 \text{ns}$ where the NMSE is always poor but is especially bad for larger $D$. This limit is separately summarized in the inset, which reveals that unless the room size is particularly small (i.e., just a few cm) high fidelity reconstruction is improssible. 

Summarizing, we see that unless very fine time resolved measurement are available, NLOS scene reconstruction becomes ill-posed and reconstruction is not robust. In the next section we show how can occluders enable high fidelity reconstruction in non-time-resolved and practical room size settings.

\section{Imaging with Occluders} \label{sec:occ_reconstruct}

The inversion problem in the poor temporal resolution limit is inherently difficult as rows of the linear forward operator $\A$ are smooth functions over the spatial target coordinate $\x$, resulting in bad-conditioning of the operator. The situation changes drastically when the line of sight between $\ellb$ (and $\cb$) and the hidden wall is partially obstructed by an occluder: for each $(\ellb,\cb)$ pair, certain segments of the hidden wall (that are different for different pairs) are occluded from $\ellb$ or from $\cb$. The occlusions are encoded in the linear forward operator $\A$ via zero entries on the corresponding spatial target coordinates $\x$, such that its rows are choppy and varied. 
Consequently, the inverse problem \eqref{eq:noisy_lin} becomes significantly better conditioned. This section builds on this idea and studies situations in which high fidelity reconstruction becomes possible even without the need for TR measurements.
%


\subsection{Informative Measurements Through Occlusions}\label{sec:occ_info}

Non-TR measurements $y_{\ellb,\cb}$ correspond to integrating \eqref{eq:fwd_model} over time, i.e., 
\begin{align} \label{eq:meas_noTR}
y_{\ellb,\cb}&= \int_{\mathcal{S}} f(\x)\frac{V(\x,\ellb)V(\x,\cb)}{\|\x-\ellb\|^2\|\x-\cb\|^2}G(\x,\ellb,\cb)\mathrm{d}\x .
\end{align}
Let $L$ be the number of distinct occluders $\mathcal{O}_i,\,i=1,\ldots,L$ that are present in the scene. We associate a distinct (binary) visibility function $V_i(\x,\z)$ to each of them. The overall visibility function $V(\x,\z)$  becomes $V(\x,\z)=\prod_i V_i(\x,\z)$, such that:
\begin{align}\label{eq:Had}
\A = \A_0\circ(\Vb_1\circ\cdots\circ\Vb_L).
\end{align}
Here, $\A_0$ is the operator corresponding to a scene with no occluders,  and $\Vb_i$ is the (binary) \emph{visibility matrix}, which has $K$ rows (as many as the number of $(\ellb,\cb)$ pairs), $N$ columns, and each of its entries takes values as follows:
\begin{align}\label{eq:vis_mat}
(\Vb_i)_{(\ellb,\cb),\x}=V_i(\x,\ellb)V_i(\x,\cb).
\end{align}
Lastly, $\circ$ denotes the Hadamard entry-wise product of matrices.
On the one hand, the operator $\A_0$ is generally badly-conditioned: successive entries of any of its rows exhibit small and smooth variations due only to the quadratic distance attenuation and the BRDF factors $G$ in \eqref{eq:meas_noTR}. On the other hand, the Hadamard multiplication with nontrivial binary visibility matrices results in a well-conditioned operator.

\begin{figure}[t!]
	\centering
	\begin{subfigure}[b]{1\columnwidth}
		\centering
		\makebox[\columnwidth]{\includegraphics[trim={2cm 0 1.5cm 0},clip,width=1\textwidth]{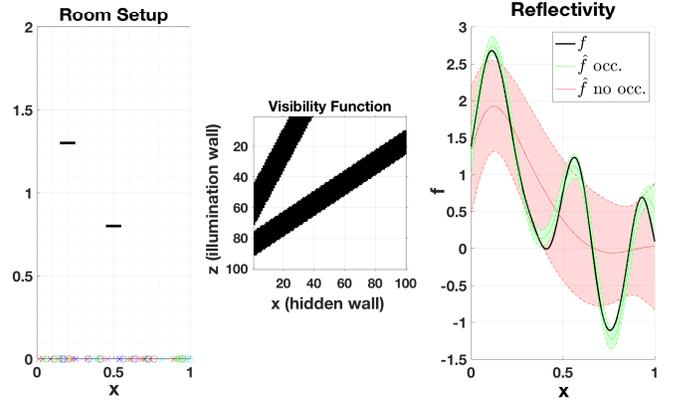}}
		\caption{(left) Room setup. On the illumination wall, positions that are marked with '$\times$' (resp. '$\circ$') indicate virtual laser (resp. camera) points.  (middle) Binary visibility matrix, with 0 (1) depicted in black (white).  (right)  Reflectivity reconstruction with (in green) and without occluder (in red).} \label{fig:occ_vs_unocc_reconstruction}
	\end{subfigure}
	\quad
	\begin{subfigure}[b]{1\columnwidth}
		\centering
		\makebox[\columnwidth]{\includegraphics[trim={2.5cm 9.4cm 2.3cm 9.5cm},clip,width=1\textwidth]{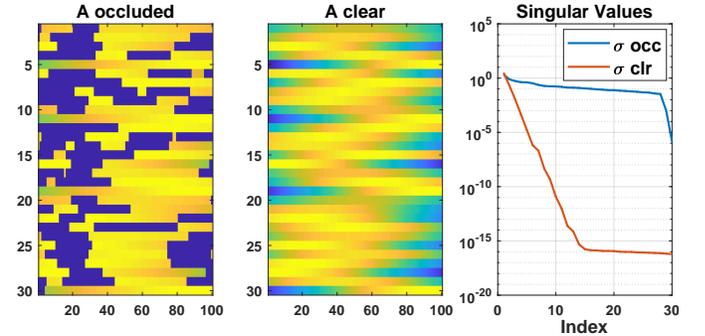}}
		\caption{ (left) Measurement matrix when occluders are present in the room. The values of its entries are depicted in Matlab's jet colormap as in Figure \ref{fig:time_resolved_reconstruction}. (middle) Measurement matrix in the absence of occluders. (right) Singular values of the two matrices in decreasing order.ֿ} \label{fig:occ_vs_unoncc_meas_mtrx}
	\end{subfigure}
	\caption{Illustrating the beneficial role of occluders, by comparing imaging in their absence and presence.}
	\label{fig:occ_vs_unoncc}
\end{figure}

This is demonstrated through an example in Figure \ref{fig:occ_vs_unoncc}, which compares reconstruction performance in the presence and absence of occluders.
The setup, illustrated in Figure \ref{fig:occ_vs_unocc_reconstruction}, is as reported in previous simulations, with the addition of occluders as depicted. We collect $K=30$ measurements with randomly drawn $\ellb,\cb$ parameters and noise variance such that $\text{SNR}=25$ dB. 
The occluded measurement matrix $\A$ and the ``un-occluded" matrix $\A_0$ are depicted in Figure \ref{fig:occ_vs_unoncc_meas_mtrx} alongside their corresponding singular values. Observe that the singular values of $\A_0$ decay substantially faster than those of $\A$, which exhibits a much flatter spectrum.
As expected, this better conditioning translates to better image reconstruction, as illustrated in the rightmost top plot: in solid red is the poor reconstruction without the occluder ($\text{NMSE}=54\%$), and in solid green is the successful reconstruction with the occluder ($\text{NMSE}=2.4\%$). The dashed lines indicate the standard deviation of the error $\hat\fb_i - \fb_i$ for each spatial coordinate $\x_i$, which corresponds to the square-root of the diagonal entries of the posterior covariance matrix. 

\subsection{Measurement Schemes}\label{sec:meas_schemes}
So far we have considered a generic setting in which a focused laser source and a focused camera generate measurements corresponding to some given set of $(\ellb,\cb)$ pairs on the illumination wall.
In principle, all possible such $\ellb$ and $\cb$ combinations are allowed. In this section we discuss the following variations to this scheme: (i) selection of most informative subset of $(\ellb,\cb)$ pairs under a budget constraint on the number of allowed measurements; (ii) measurement collection with a wide field-of-view camera; (iii) specific measurement sets that are favorable from an analysis viewpoint.

\vp
\noindent{\emph{Optimal measurement configuration:}}~
We consider a situation where collection of at most $K$ measurements is allowed, e.g., in order to limit the acquisition time of the imaging system. Under such budget constraint, we suggest an efficient strategy
to choose an optimal set $\mathcal{P}$  of $(\ellb,\cb)$ pairs and we study the imaging performance as a function of the number of allowed measurements.

Let $\Dc$ be a (uniform) discretization of the illumination wall, and $(\ellb,\cb)\in\Dc\times\Dc$.
The idea is to choose a  subset $\Sc$ such that the corresponding measurement vector $\y_\Sc:=\{\y_{\ellb,\cb}~|~(\ellb,\cb)\in\Sc\}$ is the most informative about the unknown reflectivity $\fb$. Using $I(\cdot;\cdot)$ to denote the mutual information between two (vector) random variables, this amounts to  solving
\begin{align} \label{eq:s_optimization}
\Sc^{\star}=\underset{\Sc:\Sc\subseteq \Dc\times\Dc,\vert\Sc\vert \leq K}{\text{argmax}}\; \Phi(\Sc), \quad\Phi(\Sc)
\equiv I(\y_{\Sc};\fb).
\end{align}
The optimization problem in \eqref{eq:s_optimization} is NP-hard in general. However, it turns out that under the framework of Section \ref{sec:framework}
the objective function $\Phi(\Sc)$ is monotonic and submodular (see for example \cite{shulkind2017sensor,shulkind2017experimental} for similar derivations). The theory of submodular optimization then suggests that an efficient greedy solver obtains near optimal solutions $\Sc^{\text{gr}}$ satisfying: $\Phi(\Sc^{\text{gr}})\geq (1-\frac{1}{e})\Phi(\Sc^{\star})$ \cite{fujishige2005submodular}.
The greedy algorithm augments the set $\Sc$ with an additional choice $(\ellb,\cb)$ per iteration, for a total of $K$ iterations. The solution has the property $\Sc^{\text{gr}}_{K}\subset\Sc^{\text{gr}}_{K+1}$, where we have used subscript notation for the budget constraint on the allowable size of $\Sc$. The algorithm picks the next element myopically given the solution set built so far, i.e, the algorithm picks the next element as the one which maximizes the marginal information gain.
Submodular set functions are well studied and have many desirable properties that allow for efficient minimization and maximization with approximation guarantees, e.g., \cite{fujishige2005submodular}.

\begin{figure}[t!]
    \begin{subfigure}[b]{1\columnwidth}
 	\centering
 	\includegraphics[width=0.8\columnwidth]{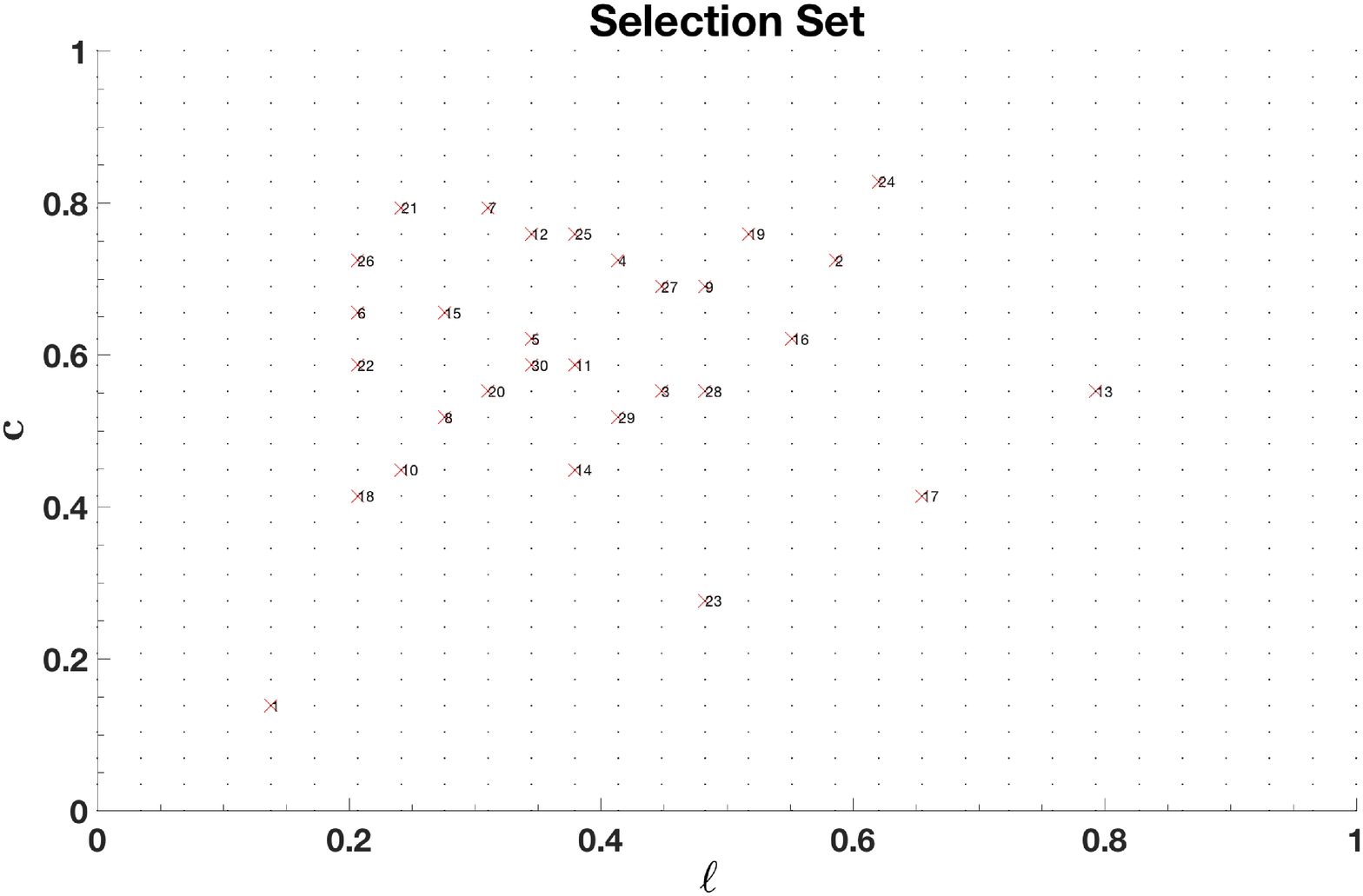}
	\caption{Coordinates of virtual laser ($\ellb$, on the horizontal axis) and camera ($\cb$, on the vertical axis) positions. The set $\mathcal{D}\times\mathcal{D}$ of all possible locations is marked with black dots. The set $\Sc$ selected by the greedy algorithm for a budget constraint $K=30$ is marked with red crosses. The numbers indicate the order of selection.}
	\label{fig:optimal_selection_greedy_output}
	\end{subfigure}
    \begin{subfigure}[b]{1\columnwidth}
  	\centering
  	\includegraphics[width=1\columnwidth]{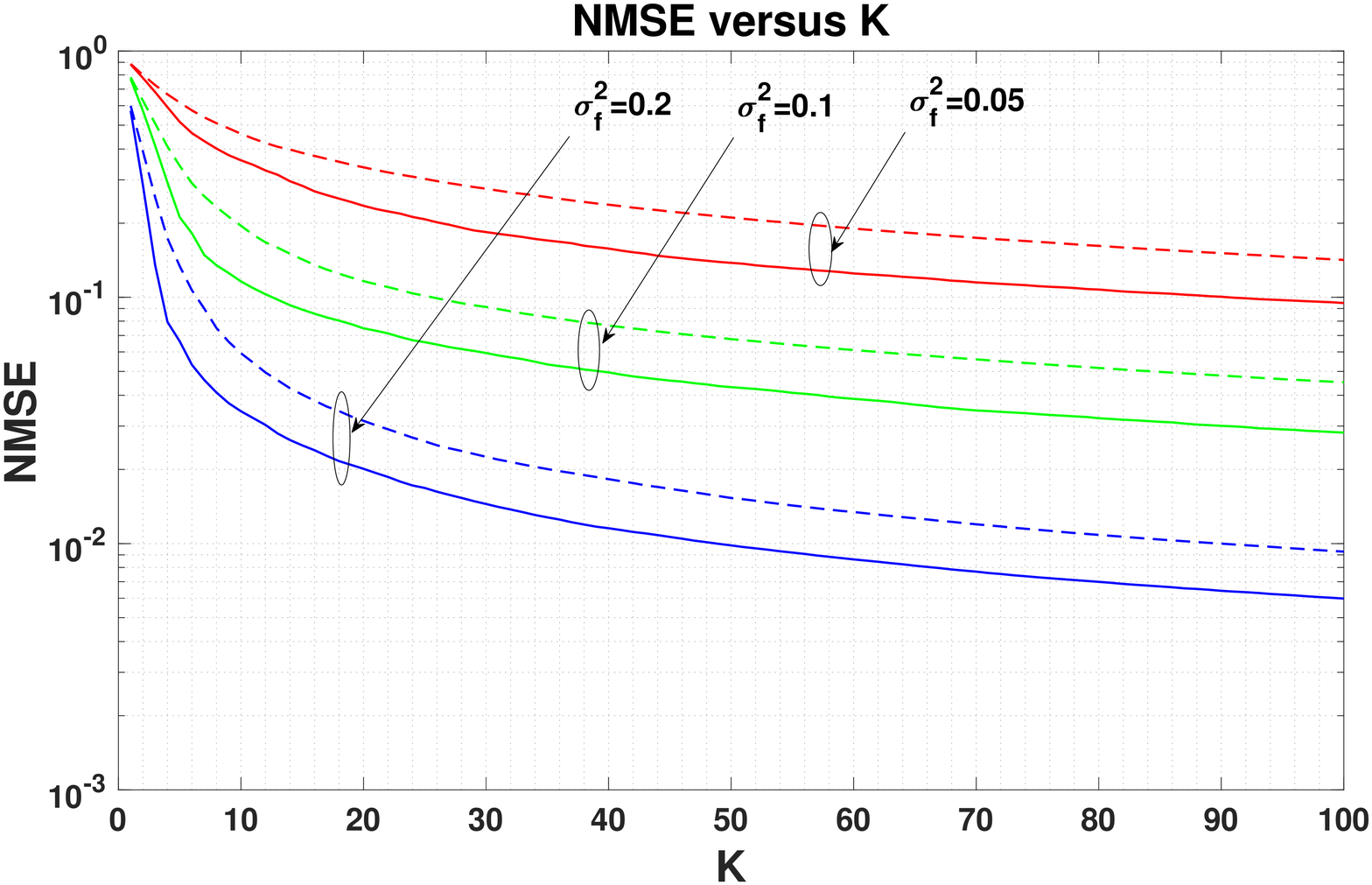}
	\caption{Reconstruction performance versus  total number of measurements for the random (dashed lines) and optimized by the greedy algorithm (solid lines) configurations for various values of the spatial correlation $\sigma_f^2$ parameter.} \label{fig:reconstruction_performance}
    \end{subfigure}
	\caption{Illustration of the efficient greedy selection algorithm for choosing informative measurements under a budget constraint.}
    \label{fig:optimal_meas}
\end{figure}

We illustrate the efficacy of this approach via numerical simulations. For the purpose of clearly illustrating the solution in a simple setting our setup is similar to that of Figure \ref{fig:occ_vs_unoncc}, except we only position one of the two occluders (the one centered around $0.5\text{m}$). The noise variance is kept constant at $\sigma^2=0.1$, and we seek an optimal set $\Sc$ of measurements under a budget constraint $|\Sc|\leq K$. Figure \ref{fig:optimal_selection_greedy_output} shows the output of the greedy algorithm for the most informative $(\ellb,\cb)$ pairs for values of $K$ up to 30. The selected parameters, marked with red crosses are accompanied by a number indicating the iteration cycle at which they were retrieved. Notice how the first two measurement configurations are selected one to the left and the other to the right of the occluder, thus casting effective shadows on different parts of the hidden wall. 
Figure \ref{fig:reconstruction_performance} validates the optimality features of the output $\Sc^{\text{gr}}$ of the greedy algorithm by comparing it to an equal size subset of measurements chosen uniformly at random. For a fixed desired NMSE the number of measurements required when randomly picking can be as large as double the number required with optimal choice. On the other hand, observe that under both schemes the NMSE drops significantly for the first few added measurements and the marginal benefit degrades as more measurements are added.

\paragraph*{Single-pixel camera with a wide field of view} 
An additional benefit from exploiting occlusion for scene reconstruction with non-TR measurements
is the ability to use a single-pixel camera with a wide field-of-view in lieu of the focused detector that is typically required for TR imaging techniques.  
This offers several advantages such as reduced equipment cost (no lens required) and a dramatically increased signal to noise ratio as more photons are collected per measurement. 
To the best of our knowledge, this is the first demonstration of NLOS imaging with a wide field-of-view detector. A camera that is configured for a wide field-of-view detects light reflected from multiple positions $\cb$ on the illumination wall; thus, it capturing more of the backscattered photons from the hidden scene. This modifies the forward measurement model as explained next.
Let $\mathcal{C}$ represent the surface of the illumination wall that is in the camera's fixed field-of-view, 
while the laser source is raster scans the illumination wall as before. This procedure yields measurements that are now parametrized only by $\ellb$, as follows:
\begin{align} \label{eq:meas_22}
&y_{\ellb}=\int_{\mathcal{C}}\frac{y_{\ellb,\cb}}{\|\cb-\boldsymbol{\Gamma}\|^2}\operatorname{cos}(\boldsymbol{\Gamma}-\cb,\mathbf{n}_{\cb}) \mathrm{d}\cb\notag \\
&= \int_{\mathcal{S}} f(\x)\frac{V(\x,\ellb)}{\|\x-\ellb\|^2}  \left[ \int_{\mathcal{C}} \frac{V(\x,\cb)G(\x,\ellb,\cb) \operatorname{cos}(\boldsymbol{\Gamma}-\cb,\mathbf{n}_{\cb})}{\|\x-\cb\|^2\|\cb-\boldsymbol{\Gamma}\|^2}\mathrm{d}\cb \right] \mathrm{d}\x.
\end{align}
In  deriving \eqref{eq:meas_22}, we used \eqref{eq:meas_noTR} and we further explicitly accounted for the quadratic power decay from the illumination wall to the position of the camera that is denoted by $\boldsymbol{\Gamma}$. The measurements are again linear in the unknown reflectivity, hence, the same reconstruction techniques can be used. In the presence of occluders, the nontrivial visibility function $V(\x,\z)$ results in a well-conditioned measurement operator and a successful image reconstruction. In particular, our experimental demonstration in Section \ref{sec:experimental_results} is based on the forward model in \eqref{eq:meas_22}.
We mention in passing that the dual setting, where a wide field-of-fiew light projector is utilized instead of a focused laser illumination, with measurements collected at multiple locations $\cb$ on the illumination wall, might also be of interest. 

\paragraph*{Other measurement configurations}
Lastly, we mention a specific configuration that reduces the dimensionality of the parameter space by imposing the restriction $\ellb=\cb$ on the measurements\footnote{Strictly speaking, when $\ellb=\cb$, the camera focused at $\cb$ sees a first-bounce response in addition to the informative third-bounce. We assume here that the dimensions of the entire scene are such that it is possible to use time-gating to reject that first-bounce. Note that this is possible with 
	mild temporal resolution requirements.}. This results in a strict subset of the entire measurement set $\Dc\times\Dc$ that is convenient for analytic purposes and for drawing insights about the features of the imaging system, and will be useful for our analysis in Section \ref{sec:features_imaging_modality}.

\subsection{Robustness} \label{sec:features_imaging_modality}

Here, we study in more detail the structural properties of the visibility function, which we use in turn to study the robustness of reconstruction with respect to a misspecified description of the location of the occluder.

\paragraph*{More on the visibility function}
Henceforth, we focus on a simple, yet insightful, case of \textit{flat
  horizontal} occluders, i.e. occluders aligned horizontally at some
fixed distance from the illumination wall (see Figure
\ref{fig:study_setup}). This family of occluders is useful as any
occluder that is small compared to the size of the room may be well
approximated as being flat and horizontal. We show that the visibility
function $V$ associated with a flat horizontal occluder has simple
structure. 
To be concrete, suppose that the occluder $\Oc$ lies on a horizontal
plane at distance $H=\alpha D$ from the visible wall for some (known)
$0<\alpha<1$. Further define the occupancy function $s(\x)$ such that
for all points $\x$ on that plane $s(\x)=0$ if $\Oc$ occupies $\x$ and
$s(\x)=1$ otherwise\footnote{Here, occluder $\Oc$ is allowed to be
  composed of several patches as long as they all lie on the same
  plane. Equivalently, the set of values for which $s(\x)=0$ need not
  be connected.}. A point $\x$ on the hidden wall is \emph{not}
visible from a point $\z$ on the illumination wall if and only if the
line that connects them intersects with the occluder, or equivalently,
if at the point of intersection it holds that
$s(\alpha\x+(1-\alpha)\z)=0$. This translates to: 
	\begin{align} \label{eq:flat_occluder_vis}
	V(\x,\z)=s(\alpha\x+(1-\alpha)\z),
	\end{align}
In particular, when $\ellb=\cb$, it follows from \eqref{eq:vis_mat} and \eqref{eq:flat_occluder_vis} that
$$(\Vb)_{(\ellb,\cb),\x}=s(\alpha\x+(1-\alpha)\ellb),$$
and the visibility function $V_i$ has a band-like structure.  Ignoring edge-effects, its discretization corresponds to a convolution matrix, which is favorable since the convolution structure makes possible deriving analytic conclusions regarding the effect of the parameters of the occluder on the image reconstruction as shown next.

\begin{figure}
\begin{subfigure}{1\columnwidth}
		\centering
		\includegraphics[width=0.39\columnwidth]{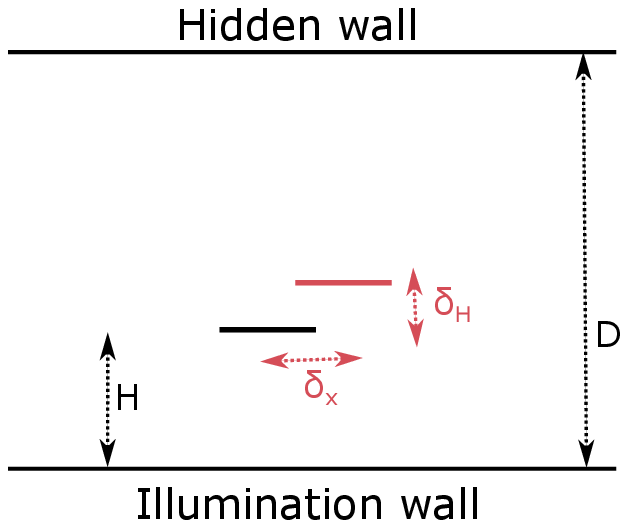}
		\caption{A shifted occluder setup. The occluder appears in its actual position in black. We perform reconstruction under imperfect knowledge of its position; taken to be as appears in red.} 
		\label{fig:occluder_shifted}
	\end{subfigure}%
	\\
	\begin{subfigure}{1\columnwidth}
		\centering
		\includegraphics[width=0.49\columnwidth]{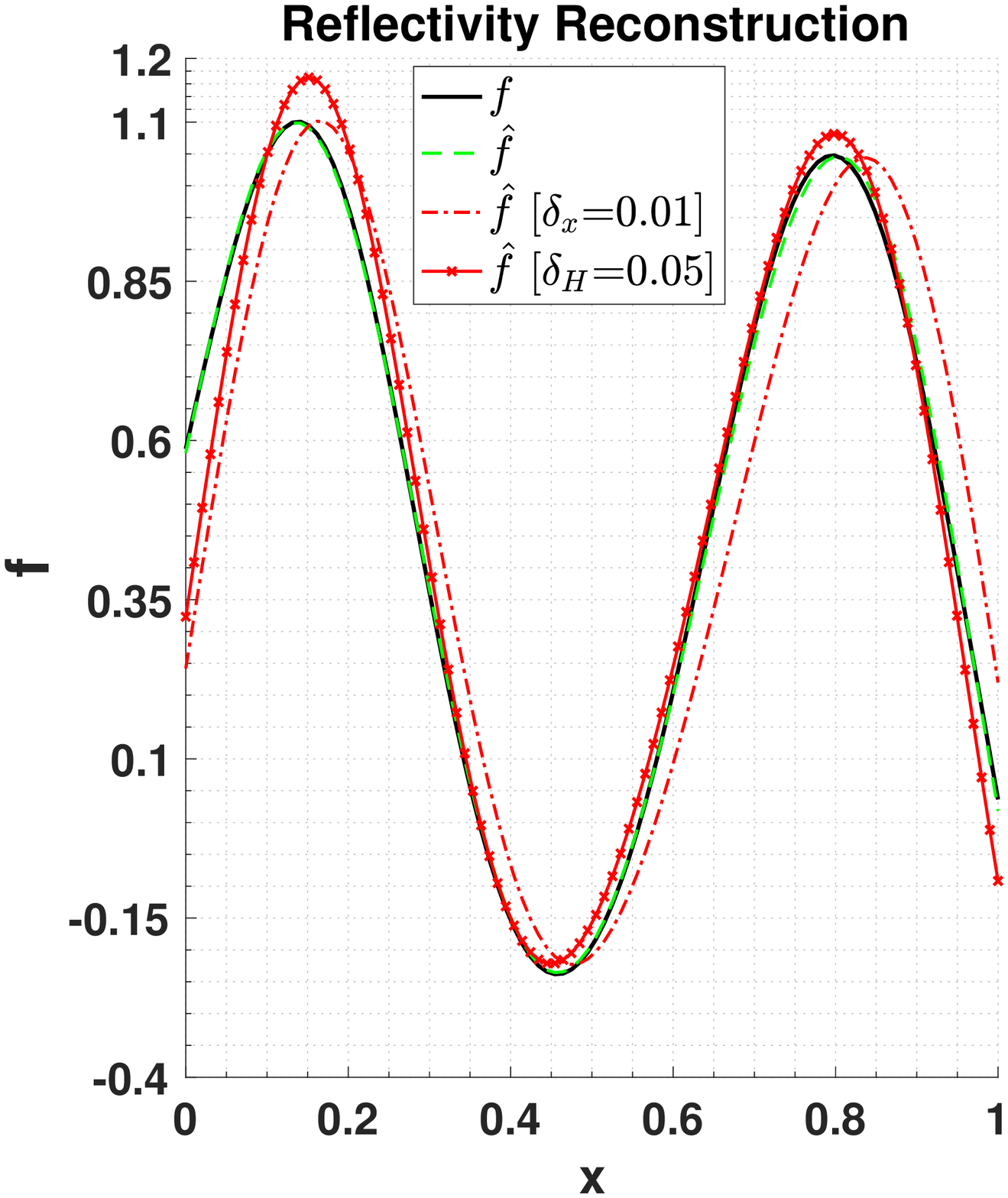}
		\includegraphics[width=0.49\columnwidth]{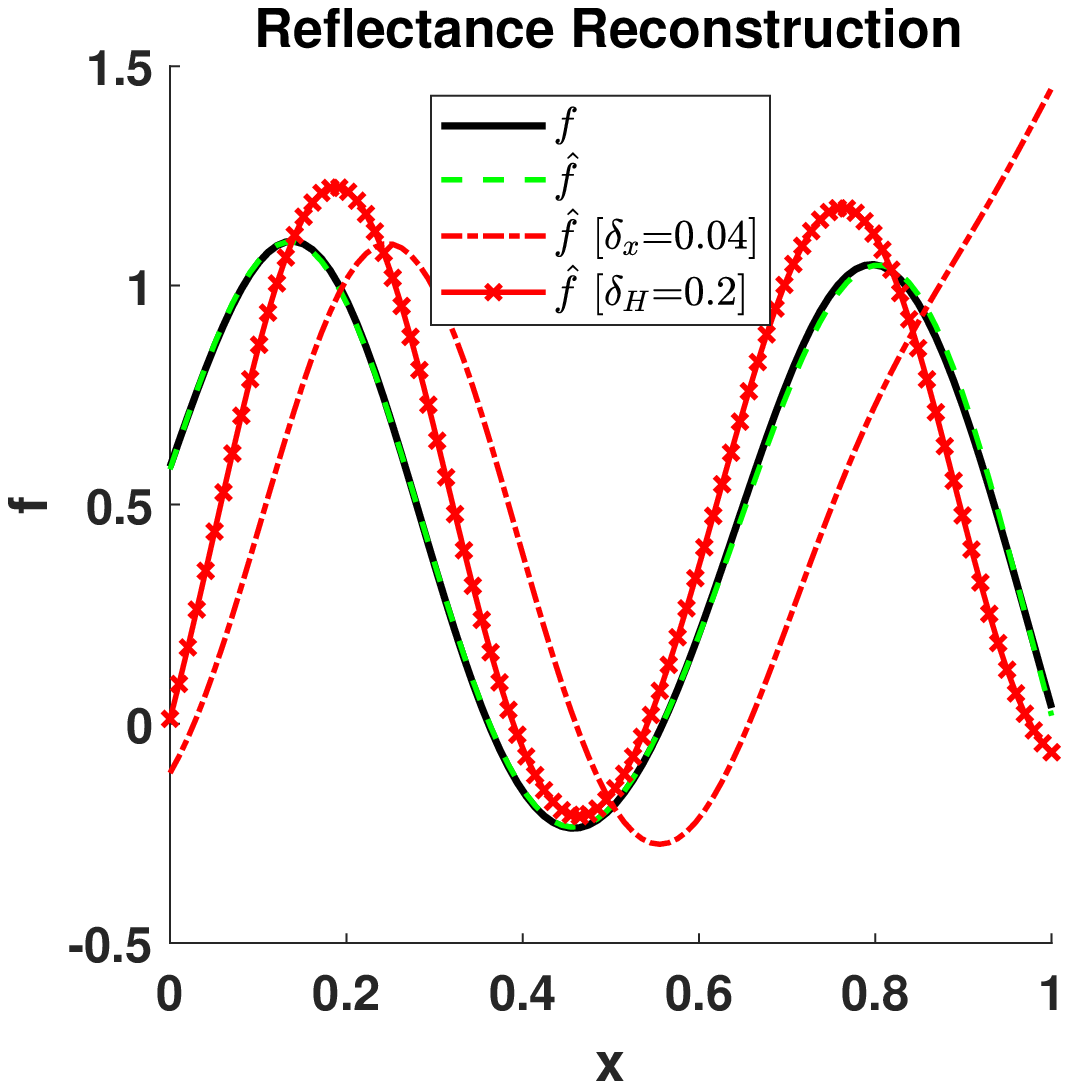}
		\caption{Reconstruction with a mispositioned occluder. (left) Small  and (right) large  vertical and horizontal shifts in a far field setup.}
	\end{subfigure}%
	\caption{Illustrating the effect of modeling mismatches  on reconstruction.}
		\label{fig:robust_rec}
\end{figure}

\paragraph*{The effect of modeling mismatches}
We study scene reconstruction under a mismatched model for the
position of the occluders, to evaluate the robustness of our imaging
method with respect to such modelling errors. Figure
\ref{fig:occluder_shifted} illustrates our setup where the true
position of the occluder appears in black, and our mismatched model
assumes the occluder is positioned as appears in red, with $\delta_x$
and $\delta_H$ vertical and horizontal shifts, respectively. We study
the resulting reconstruction under the following simplifications: (i)
measurements are noiseless, (ii)  measurements are taken with
parameters satisfying $\ellb=\cb$, (iii) continuous measurements are
collected, i.e. $y_{\ellb}$ is available for all points $\ellb$ on the
visible wall, and (iv) we assume that the hidden wall is far from the
illumination wall such that $\|\x-\ellb \|^2\|\x-\cb\|^2$ and
$G(\x,\ellb,\cb)$ are approximately constant. 

Under these assumptions, the measurements $\y_{\ellb}$ are expressed
(up to a constant) as  
\begin{align} \label{eq:fwd_model_no_error}
&\y_{\ellb} = \int f(\x) s(\alpha\x+(1-\alpha)\ellb){\rm d}\x,
\end{align}
where we have used \eqref{eq:flat_occluder_vis},  and $f(\x)$ is the true reflectivity of the hidden wall. 

In the presence of errors $\delta_x,\delta_H$,  the misspecified visibility function can be expressed as
$\tilde{V}(\x,\z)=s(\alpha'(\x-\delta_x)+(1-\alpha')(\ellb-\delta_x)),$
where $\alpha':=\frac{H+\delta_H}{D}=\alpha+\frac{\delta_H}{D}$. This results in a misspecified model:
\begin{align} \label{eq:fwd_model_error}
&{\y}_{\ellb} = \int \hat{f}(\x) s(\alpha'(\x-\delta_x)+(1-\alpha')(\ellb-\delta_x)){\rm d}\x.
\end{align}

In order to study how $\hat f(\x)$ relates to $f(\x)$ it is convenient to work in the Fourier domain\footnote{\label{foot:Fourier}
	The variable of integration $\x$ in \eqref{eq:fwd_model_no_error} and \eqref{eq:fwd_model_error} ranges over the finite surface of the hidden wall. Correspondingly, $f(\x)$ and $s(\x)$ are only defined over this region. Formally, when it comes to taking Fourier transforms, we extend the functions on the rest of the space by zero-padding.}. 
Taking Fourier transforms of the right-hand-side expressions of both \eqref{eq:fwd_model_no_error} and \eqref{eq:fwd_model_error}, and equating, it can be shown that\footnote{Recall $\mathcal{F}[f(t)]=F(\omegab)\rightarrow\mathcal{F}[f(at+b)]=\frac{1}{\vert a\vert}e^{-j\omegab\frac{b}{a}}F(\frac{\omegab}{a}).$}
\begin{align}\label{eq:FT}
{\hat F}(\omegab)=\frac{1-\alpha'}{1-\alpha}\frac{S(-\frac{1-\alpha'}{1-\alpha}\frac{\omegab}{\alpha'})}{S(-\frac{\omegab}{\alpha'})}e^{j\omegab\frac{\delta_x}{\alpha'}}F\left(\frac{\alpha}{\alpha'}\frac{1-\alpha'}{1-\alpha}\omegab\right),
\end{align}
where $G(\omegab)$ denotes the Fourier transform of a function $g(\x)$. Of course, this holds for frequencies at which $S(\omegab)$ is non-vanishing.

The following conclusions regarding  reconstruction distortion under mismatched occluder position 
are drawn from \eqref{eq:FT}. (a) In the absence of errors ($\delta_x,\delta_H=0$), the reflectivity function is perfectly reconstructed for those frequencies for which the shape-function of the occluder is non-zero. (b) Horizontal occluder translation errors ($\delta_x\neq0, \delta_H=0$) result in simple shifts of the true reflectivity. (c) Vertical occluder translation errors ($\delta_x=0, \delta_H\neq0$) result in two kinds of distortion. The first is a scaling effect, while the other is a distortion that depends on the shape of the occluder through the term $S(-\frac{1-\alpha'}{1-\alpha}\frac{\omegab}{\alpha'})/
S(-\frac{\omegab}{\alpha'})$. For this latter term, observe that its effect diminishes for a  spectrum $S(\omegab)$ that is mostly flat over a large range of spatial frequencies.  This property is approximately (due to the finite support$^{\ref{foot:Fourier}}$ of $s(x)$) satisfied by a very narrow occluder.

Recall that the above conclusions hold analytically in the limit of a distant hidden scene and a continuum of noiseless measurements. However, the conclusions are also suggestive and insightful for practical scenarios, as illustrated by the numerical study shown in Figure \ref{fig:robust_rec}, where we illustrate high SNR ($35 \text{dB}$) reconstruction with a mispositioned occluder. The room setup is $D=5$m, with a single occluder of width $0.25$m positioned at $[.5,2]$m. 
Measurements are collected with random $\ellb$ and random $\cb\neq\ellb$. Black solid lines show the true reflectivity $f(\x)$. The dashed green line depicts reconstruction under perfect occluder knowledge. The red curves show reconstructions with horizontally and vertically mispositioned occluders. The mispositioning is larger in the right subplot. It is evident from the images that horizontal mispositioning mostly results in a shifted reconstruction, whereas vertical mispositioning results in axis-scaling of the reconstructed scene. Our analysis-based conclusions seem to be valid for the middle section of the reflectivity function, whereas edge effects appearing close to the boundaries $x=0,1$ are not captured by the analysis.

The robustness of our imaging method with respect to occluder positioning errors is further supported by the  experimental demonstration in Section \ref{sec:experimental_results}, where such occluder modelling inaccuracies are unavoidable, yet the reconstruction results we demonstrate are satisfactory. 

\subsection{Reconstruction of Reflectivity with Unknown Distance} \label{sec:joint}
Thus far, we have demonstrated the use of occluders to reconstruct the unknown reflectivity of a hidden wall when its geometry is known. 
Here, we develop a simple algorithm for reflectivity reconstruction with the aid of occluders when the \emph{distance} $D$ between the visible and hidden walls is unknown.
%


In line with the Bayesian approach in Section \ref{sec:bayes}, we associate some distribution with the unknown depth $D$, and attempt joint estimation of both $D$ and $\fb$ by solving the maximum a posteriori (MAP) problem:
\begin{align}\label{eq:MAP_1}
(\hat{D},\hat{\fb}) = \arg\max_{D',\fb'}~ p(D',\fb'|\y),
\end{align}
where $\y=\A_D\fb+\mathbf{\epsilon}$ as in \eqref{eq:noisy_lin}. Observe that the distance $D$ enters the measurement equations via the forward operator, which we have \cmmnt{denoted} parametrized as $\A_D$. For a fixed $D'$ the maximization in \eqref{eq:MAP_1} with respect to $\fb$ has already been studied in terms of (efficient) implementation and performance. Namely, under a Gaussian prior assumption on $\fb$, each maximizer $\fbh_{D_i}$ coincides with the MMSE estimator of Sec.~\ref{sec:bayes}. Based on this observation, a simple and effective strategy for solving the joint optimization in \eqref{eq:MAP_1} is as follows. Start with a range of candidate distance values, $D_1,D_2,\ldots,D_N$. For each candidate, form the measurement matrix $\A_{D_i}$ and solve for the corresponding reflectivity vector $\fbh_{D_i}$. Then, for $i=1,2,\ldots,N$, compute $i_*$ that maximizes the likelihood (we assume here a uniform prior among the $D_i$'s):
$$
i_* = \arg\max_i p(\y | \fbh_{D_i}, D_i).
$$
Finally, return $(\hat{D},\fbh)=(D_{i_*},\fbh_{i_*})$. 
In particular, under the Guassian prior assumption, it can be shown that 
\begin{align}\label{eq:logP_Gauss}
-\log p(\y | \fbh_{D_i}, D_i) = \y^\top(\A_{D_i}\Sigma_\fb\A_{D_i}^\top+\sigma^2\mathbf{I})^{-1}\y.
\end{align}
Note however that the algorithm can be readily adapted to different priors on $\fb$.

\begin{figure}
\begin{subfigure}{1\columnwidth}
		\centering
		\includegraphics[width=1\columnwidth]{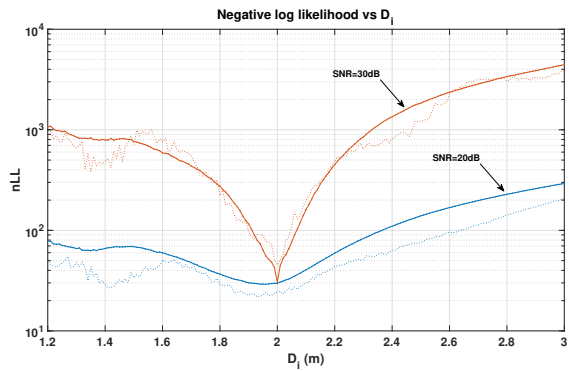}
		\caption{Plot of the negative log-likelihood (nLL) function versus each candidate distance value $D_i$ (see Equation~\eqref{eq:logP_Gauss}) for two different values of the SNR and for $K=30$ measurements. The solid lines represent averages over 200 realizations of the reflectivity and of the measurement positions. The dashed lines show the nLL for a specific such realization. } \label{fig:joint_nLL}
	\end{subfigure}%
	\\
	\begin{subfigure}{1\columnwidth}
		\centering
		\includegraphics[width=1\textwidth]{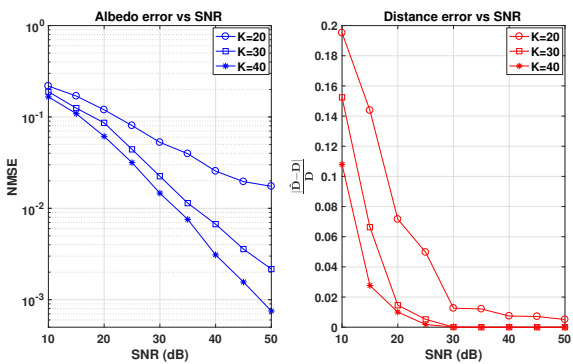}
		\caption{Plots of (normalized) reconstruction error for the reflectivity and the distance as a function of the SNR for different number of measurements.  }	\label{fig:joint_err}
	\end{subfigure}%
	\caption{Illustration of the proposed algorithm for reflectivity estimation when the distance $D$ is uknown.}
	\label{fig:joint}
\end{figure}

Figure \ref{fig:joint} includes an illustration of the algorithm and a numerical demonstration of its performance for different values of parameters such as SNR and number of measurements. The room setup is the same as in Fig. \ref{fig:occ_vs_unoncc}. In particular, the true distance of the hidden wall is $D=2$ and the reflectivity is drawn from a Gaussian prior with $\sigma_f^2=0.05$. A total number of $K$ randomly selected $(\ellb,\cb)$-measurements are collected. Observe in Figure \ref{fig:joint_nLL}  that the negative log-likelihood in \eqref{eq:logP_Gauss} shows a valley in the neighborhood of the true distance $D$. Higher values of SNR result in sharper valleys and the minimum occurs at the true distance (here $D=2$) provided that enough measurements are available (see Fig. \ref{fig:joint_err}). The plots shown are averages over 200 realizations drawn from the Gaussian prior with each instance measured by 30 randomly selected $(\ellb,\cb)$ pairs. 

\begin{figure}[h]
	\centering{\includegraphics[width=1\columnwidth]{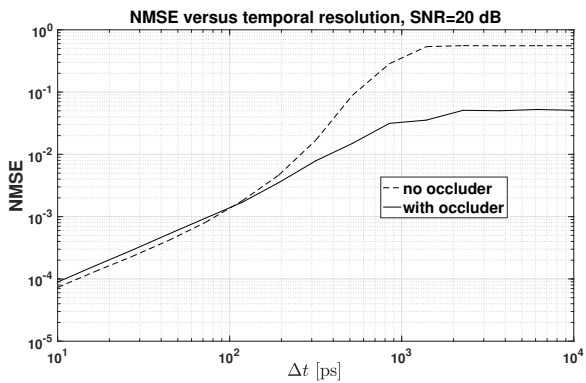}}
	\caption{Comparing reconstruction performance vs. temporal resolution in the presence and absence of occluders.} \label{fig:se_vs_occ}
\end{figure}

\subsection{Collecting TR-Measurements in Occluded Settings} \label{sec:TR_vs_occ}
Thus far we have focused on imaging systems that use either  TR measurements or non-TR measurements and occlusions. It is natural to attempt combining the best of both worlds. A full study of this topic is beyond the scope of the paper, but we present numerical simulations to illustrate its promises.
%
Consider the familiar setting of Figure \ref{fig:occ_vs_unocc_reconstruction} and a detector with a nontrivial temporal resolution $\Delta t$. We sweep $\Delta t$ over a range of values, and plot the resulting NMSE in Figure \ref{fig:se_vs_occ} (solid curve). For comparison, we also plot in dashed line the NMSE performance in the absence of an occluder (this corresponds exactly to the plot in Figure \ref{fig:mse_vs_res_vs_SNR}). For a large range of temporal resolutions (here, $\Delta t \gtrsim 150 \text{ps}$) the presence of occlusions leads to a substantial improvement in reconstruction performance, allowing the same level of performance to be maintained at inferior temporal resolution levels. When very high temporal resolution is available, the reconstruction performance is almost identical whether occluders are present or not.
Note here that TR measurements can be further utilized to improve other aspects of the system. For instance, one might imagine using (coarse) TR measurements to find the position of the occluder more effectively than could otherwise be possible. We comment more on this in Section \ref{sec:discussion}. 

%
\section{Experimental Illustration} \label{sec:experimental_results}
We experimentally demonstrate an instance of opportunistic exploitation of occluders to perform NLOS active imaging with non-TR measurements. An extension of the methods to the low-photon count regime and additional results can be found in \cite{experiment}.

\begin{figure}
	\centering{\includegraphics[width=1\columnwidth]{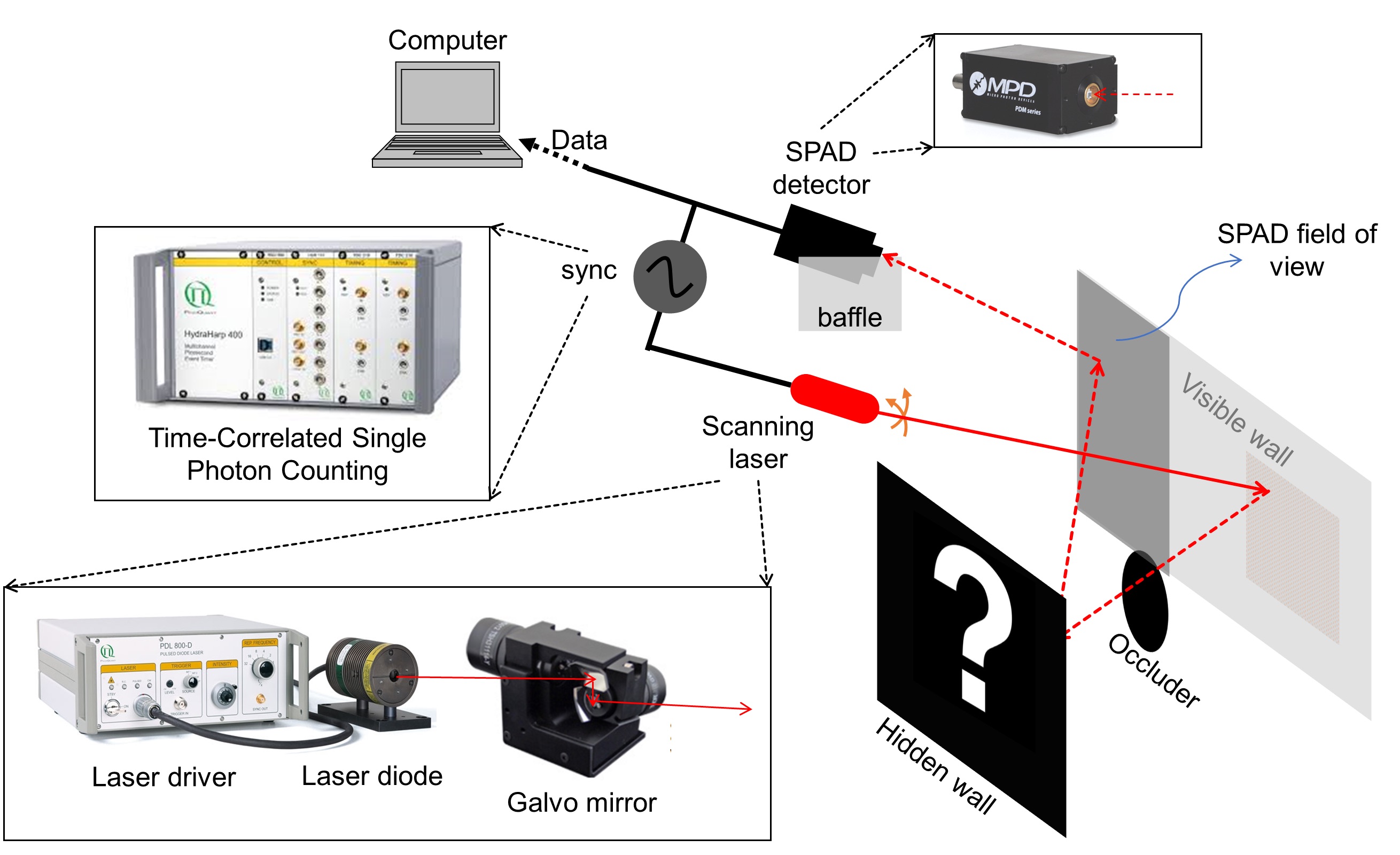}}
	\caption{The experimental setup, with the following distances - visible wall to hidden wall: $\sim$106 cm; visible wall to SPAD: $\sim$156 cm; visible wall to occluder: 37 cm; The diameter of the circular occluder is 3.4 cm.}
	\label{fig:experimental_room_setup}
\end{figure}

\paragraph*{Experimental Setup}
The schematic setup of our experiment is shown in Figure
\ref{fig:experimental_room_setup}. A pulsed 640-nm laser source
illuminates a nearly Lambertian-surface visible wall (1st bounce). The
forward-going light travels to the hidden wall, which scatters the
light back (2nd bounce). Finally, the backscattered light from the
visible wall (3rd bounce) is collected by a SPAD detector\footnote{A
  SPAD is capable of providing time-resolved information. However, for
  the purpose of this experiment we operate the SPAD as a regular
  camera, essentially integrating the response over time. To be
  precise, we only use the time resolved measurements of the SPAD to
  gate-out the first-bounce response from the illumination
  wall. Beyond that, no TR measurements are recorded. Notice that the
  illumination wall is in the direct line of sight of the imaging
  equipment, thus its location can be well-estimated based on standard
  imaging techniques. With this information, the time window that
  corresponds to the first-bounce response is a-priori known. Hence,
  the same operation achieved here with a SPAD camera can be perfomed
  using a CCD camera.}. In front of the SPAD, an interference filter
centered at 640 nm is used to remove most of the background light. In
the experiment, the SPAD is lensless and configured for a
wide-field-of-view detection of the left side of the visible wall to
minimize 1st bounce light detection. The occluder is a black-surface
circular patch without any back reflections. During the experiment, we
turned off all ambient room light to minimize background noise. 

\begin{figure}
	\centering{\includegraphics[width=1\columnwidth]{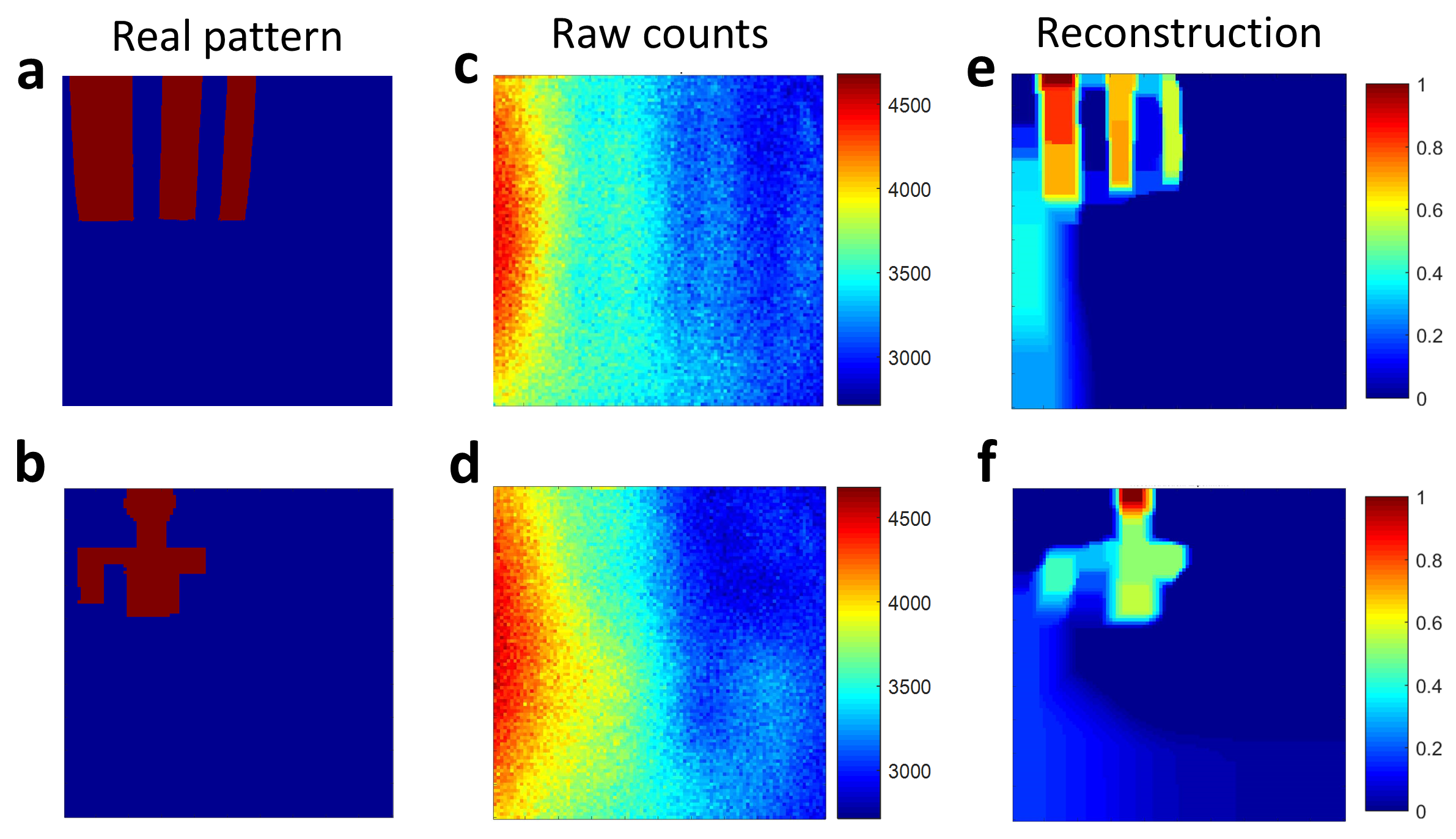}}
	\caption{(\textbf{a,b}) Ground truth of the tested scene patterns on the hidden wall. The patterns are placed in the upper-left corner of the hidden wall. (\textbf{c,d}) The raw measurement counts for a $100\times 100$ raster-scanning laser points. At each laser point, we turn on the SPAD for a fixed dwell time such that $\sim$3500 photon counts are recorded on average. (\textbf{e,f}) Reconstruction results from Eq.~\eqref{eq:exp_TV}.} \label{fig:exp_reconstruct1}
\end{figure}

\begin{figure}
	\centering
	\includegraphics[width=1\columnwidth]{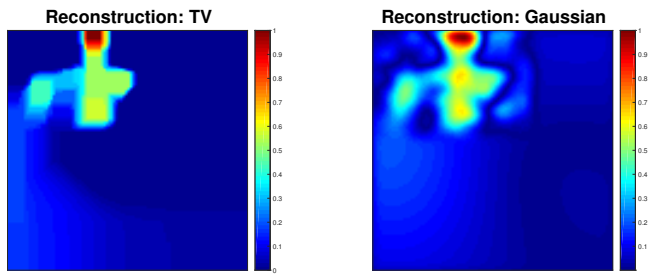}
	\caption{Comparing reconstruction results with the nonlinear TV-based method in \eqref{eq:exp_TV} (left) to ones with the linear method in \eqref{eq:f_hat} that is based on the Gaussian prior model (right).} \label{fig:exp_reconstruct2}
\end{figure}

\paragraph*{Computational processing}
Based on the forward model in Equation~\eqref{eq:meas_22} we 
obtain an estimate $\mathbf{\hat{f}}$ of the true reflectivity by solving the following non-smooth convex optimization program:
\begin{align}\label{eq:exp_TV}
\mathbf{\hat{f}} = \arg\min_\fb \frac{1}{2}\|\y-\A\fb\|_2^2 + \la\|\fb\|_{\text{TV}},
\end{align}
where $\|\cdot\|_\text{TV}$ is the Total-Variation (TV)-norm and $\la>0$ is a regularization parameter.  To solve \eqref{eq:exp_TV} we use an efficient dedicated iterative first-order solver~\cite{harmany2012spiral}, which is based on the popular FISTA algorithm \cite{beck2009fast}. TV-norm penalization is a standard technique that has been successfully applied in other imaging tasks (e.g., image restoration \cite{levin2007image,beck2009fast,krishnan2009fast}). Its use is motivated by the observation that the derivatives of natural images obey heavy-tailed prior distributions \cite{levin2007image,krishnan2009fast}. In Figure \ref{fig:exp_reconstruct2} we compare the nonlinear TV-based reconstruction to the linear reconstruction in \eqref{eq:f_hat} that assumes a Gaussian prior on $f(\x)$ (see Section \ref{sec:bayes}) with $\sigma_f^2=0.02$ and $\sigma^2$ tuned to achieve good results.
TV regularization is more accurate and emphasizes edges as expected. The linear one is blurry but satisfactory and yields a reconstruction that is easily interpretable by human eye. One should also note that the linear reconstruction is more computationally efficient. Both methods require tuning of the involved parameters: $\sigma_f^2$ and $\sigma^2$ for GP, and $\lambda$ for TV.

In \cite{experiment}, we appropriately specialize the reconstruction algorithm \eqref{eq:exp_TV} to account for the Poisson statistics of the measurements  in low-photon count regimes. Specifically, we substitute the least-squares term in \eqref{eq:exp_TV} with the negative log-likelihood from the Poisson statistics. In the current demonstration we allowed for long exposition times and high-photon measurements, in which case it is seen that the performance of \eqref{eq:exp_TV} is satisfactory.

\paragraph*{Results} Reconstruction results using the optimization
method in \eqref{eq:exp_TV} are shown in
Figure~\ref{fig:exp_reconstruct1}. The regularizer parameter $\lambda$
was tuned independently for each algorithm to yield a reconstruction
that is empirically closest to the ground truth. Tuning in this manner
is convenient for such demonstrations, but in the absence of truth,
one typically resorts to a cross-validation procedure. In Figure
\ref{fig:exp_reconstruct1}, two different scene patterns on the hidden
wall were tested. The laser light was raster scanned on a $100\times
100$ grid and at each point, the SPAD detector was turned on for a
fixed dwell time such that a total number of $\sim$9 million laser
pulses ware emitted and $\sim$3500 back-reflected 3rd bounce photons
were recorded on average. The raw measurement counts for each of the
hidden patterns are shown in Figures ~\ref{fig:exp_reconstruct1}(c,d):
each one of the $100\times 100$ entries corresponds to a measurement
$\y_{\ell}$ for the corresponding virtual laser position $\ellb$. The
raw counts are processed by the optimization algorithm in
\eqref{eq:exp_TV} to obtain an estimate of the hidden patterns as
shown in Figures~\ref{fig:exp_reconstruct1}(e,f). 
These results validate the forward model and the performance of the
reconstruction algorithm. 

\section{Discussion and future work} \label{sec:discussion}
In this paper we introduce and explore the benefits of exploiting occlusions in NLOS imaging. 
We focus on the problem of reconstructing the reflectivity of a hidden surface of known geometry from diffuse reflections, further assuming that the occluders in this setup are %
absorbing and of known geometry. 
This serves as a useful testing ground for demonstrating basic principles in occluder-assisted NLOS imaging. At the same time, our promising results suggest that it is of interest to extend the study to more complicated system models. It further suggests exploring the premises of opportunistic NLOS imaging under even broader settings. In what follows, we elaborate on  relevant directions of future research. 
%

Beyond the problem of reflectivity estimation, it 
remains to
explore extensions towards full 3D reconstruction of more complicated scenes. 
While much of our focus has been on identifying scenarios where the use of occluders can alleviate the need to collect TR measurements, we speculate that combined use of both TR measurements and occluders can assist in approaching more complicated problems such as the aforementioned.

Another interesting extension is as follows.
Rather than using known occluders to reconstruct the reflectivity function, one can imagine scenarios where the reflectivity function of a back wall is known, thus it can be exploited to identify the the position of unknown objects in the hidden room.
In terms of the forward model in \eqref{eq:fwd_model} this essentially asks for an estimate of the visibility function given the measurements and the reflectivity $f(\x)$, since the visibility function is in turn informative about the shape of the occluders. 

Continuing along the same lines it is natural to consider the fully blind problem, in which both the reflectivity function and the occluder shape are unknown. A natural approach to solving this problem is an iterative alternating-optimization method, which iterates between the two subproblems that were previously discussed: solve for $f(\x)$ given the shape function, and vice-versa. For each subproblem, we can use convex-optimization with appropriate regularization to promote the statistical or structural properties of the desired quantities (e.g., Gaussian prior on $f(\x)$ and a low total-variation assumption on the shape function). Analyzing the convergence properties of such procedures and further understanding the extent to which different priors are sufficient to identify the true underlying quantities are compelling research questions. 

Similar to the use of occluders as a form of opportunistic imaging, it is possible that  exploiting other structural features of the environment results in enhancement of  NLOS imaging.
As already discussed, one such example involves exploiting the possibly known reflectivity pattern in back walls. Another example is about utilizing coincidental bumps or edges on the illumination wall itself and the occlusions that those introduce. 
Finally, it is natural to attempt extensions of the discussed methods to non-static environments.
For instance, moving occluders will generate measurements with additional diversity that can be exploited towards more accurate and robust reconstructions. 

\bibliographystyle{ieeetr}
\bibliography{REVEAL}
\end{document}